\documentclass[aps,prd,amssymb,amsmath,nobibnotes,nofootinbib,superscriptaddress, reprint, 12pt ]{revtex4}
\usepackage{amsfonts,amsmath,hyperref,url, color}
\allowdisplaybreaks
\usepackage{bm, bbm}
\usepackage{graphicx}
\usepackage{caption}
\usepackage{mathtools}
\usepackage{t1enc}
\usepackage{hepnames}
\usepackage{tensor}
\usepackage{comment}
\usepackage{float}
\usepackage{wrapfig}

\usepackage{dcolumn}

\usepackage{adjustbox,booktabs}

\usepackage{makecell}
\usepackage{array}
\DeclareMathAlphabet\mathbfcal{OMS}{cmsy}{b}{n}

\usepackage{soul}

\newcommand{\bc}{\begin{center}}
\newcommand{\ec}{\end{center}}
\newcommand{\be}{\begin{eqnarray}}
\newcommand{\ee}{\end{eqnarray}}
\newcommand{\bs}{\begin{slide}}
\newcommand{\es}{\end{slide}}
\newcommand{\la}{\langle}
\newcommand{\ra}{\rangle}
\newcommand{\bi}{\begin{itemize}}
\newcommand{\ei}{\end{itemize}}
\newcommand{\nn}{\nonumber}

\begin{document}

\title{Quantum Gravity Phenomenology and Particle Physics}

\author{Andrea Bevilacqua}
\affiliation{National Centre for Nuclear Research, ul. Pasteura 7, 02-093 Warsaw, Poland}
\author{Jerzy Kowalski-Glikman}
\affiliation{University of Wroc\l{}aw, Faculty of Physics and Astronomy, pl.\ M.\ Borna 9, 50-204
Wroc\l{}aw, Poland}
\affiliation{National Centre for Nuclear Research, ul. Pasteura 7, 02-093 Warsaw, Poland}
\author{Wojciech Wi\'slicki}
\affiliation{National Centre for Nuclear Research, ul. Pasteura 7, 02-093 Warsaw, Poland}

\date{\today}

\begin{abstract}

Quantum gravity  phenomenology has been historically regarded as a difficult endeavour, due to the apparent scarcity of phenomena involving the required scales of length (Planck length $l_P$) and energy (Planck energy $E_P$). It was realized, however, that one can look for cumulative effects of a quantum theory of gravity  at energies $E/E_P \ll 1$ if even tiny effects are amplified by the large time of flight or very high energies, common to astrophysical phenomena. 

In the this work, complementary to the COST action CA18108 White Book, we put forward a proposal that quantum gravity phenomenology can be fruitfully pursued also with the help of terrestrial particles accelerators. We first discuss the theoretical background, and  then concentrate on deformed discrete symmetries C,P,T. We investigate possible experimental signatures of CPT deformation, particularly concerning the difference in decay time between particles and antiparticles, and fields interference.

\end{abstract}

\maketitle

\section{Introduction}

Ever since the dawn of quantum gravity research project \cite{Bronstein:2012zz} almost 90 years ago, it was expected that the progress in the field was going to be slow as a result of the very sparse experimental feedback. Indeed, to scatter elementary particles at Planckian energies ($\sim 10^{19} $ GeV) with Planck length impact parameter ($\sim 10^{-35} $ m) would require particle accelerator of cosmic size, clearly unimaginable even in the distant future. This fact has led some of the researchers to claim \cite{Dyson:2013hbl} that the direct quantum gravity scattering processes might not be detectable, even in principle.

It  came as a kind of surprise to the community when about 25 years ago it was argued \cite{Amelino-Camelia:1999hpv} that quantum gravity phenomenology is not only possible in the near future, but some of the tests can be even performed with the help of current technologies (see Refs. \cite{Amelino-Camelia:2008aez}, \cite{Addazi:2021xuf} for recent reviews). The game changing observation was that in order to get an insight into the structure of quantum gravity theory, one does not need to observe the Planck scale scattering; instead one may try to look for possible imprints of quantum gravity on (relatively) low-energy phenomena in physical regimes, in which the minute quantum gravity effects are somehow amplified.

The key element on which virtually all the approaches to the quantum gravity agree is that the nature of quantum spacetime becomes radically different from the standard smooth pseudo-Riemannian manifold of general relativity. The so called ``spacetime foam'' structure comes forth when one  starts probing distances close to the Planck scale (see e.g. Ref. \cite{Hossenfelder:2012jw}). In the emerging picture  of a ``quantum spacetime''  the very concept of spacetime points, on which all our standard theories are based, inevitably breaks down at the Planck scale, as a result of a fundamental limitation to the measurability of distances. These considerations have led to the claim \cite{Doplicher:1994zv}, \cite{Doplicher:1994tu}, \cite{Majid:1999tc} that the non-commutative geometry is the appropriate tool to describe physics at the Planck scale. The particularly promising model of this kind, which already has proved its phenomenological potential in the astrophysical context \cite{Addazi:2021xuf},  is $\kappa$-deformation.

The $\kappa$-deformation was first formulated (under the name of the $\kappa$-Poincar\'e algebra) in early 1990s \cite{Lukierski:1991pn,Lukierski:1991ff,Lukierski:1992dt,Lukierski:1993wx,Majid:1994cy} (see \cite{Arzano:2021scz} for review). It later became the working example of Doubly Special Relativity \cite{Amelino-Camelia:2000cpa}, \cite{Amelino-Camelia:2000stu} and Relative Locality \cite{Amelino-Camelia:2011lvm} scenarios. The main idea here was the emergence of  a semiclassical regime of quantum gravity characterized by the presence of a scale $\kappa$ of dimension of mass, identified with Planck scale. By the correspondence principle $\kappa$-deformation disappears in the limit $\kappa\hookrightarrow\infty$. In this limit $\kappa$-deformed theory (of particles and/or fields) reduces to its standard relativistic counterparts. It is important from a phenomenological perspective that the corrections to well-established theories, in the leading order generically behave like $\mathcal{E}/\kappa$, where $\mathcal{E}$ is the energy scale of the process in question. In the context of the current accelerator technologies these corrections are therefore of order of $10^{-15}$ and can be improved by one or two orders of magnitude in  future accelerators. The quantum gravity induced corrections seem to be very small, but fortunately large enough to be in principle measurable with the help of the very sensitive, dedicated experimental setup.

So far, investigations in quantum gravity phenomenology were mostly focused on the analyses of high energy particles created by distant astrophysical sources and then detected by terrestrial and/or satellite observatories. Assume, for example (see Refs. \cite{Amelino-Camelia:1997ieq} and \cite{Addazi:2021xuf} for detailed discussion), that the vacuum dispersion relation of massless particles is modified by quantum gravity, so that photons of different energies move with different speeds $\Delta v = \Delta E/\kappa$, where $\kappa$ is the quantum gravity energy scale. If we have a very distant source of a wide spectrum of photons, the difference in the arrival time of  photons of different energies  is a product of speed difference and the distance from the source. In the case of a source at a cosmological distance from Earth, of order of billions light years, with the help of such observations we can put the bound of the scale $\kappa$ up to the Planck scale.
Investigations of this kind of effects was the main focus of the recent COST action CA18108 ``Quantum Gravity Phenomenology In The Multi-Messenger
Approach'' \cite{cost}. Such observations have clear advantages. The astrophysical sources are capable of producing particles carrying energies many orders of magnitude larger than the ones accessible in accelerators. Moreover, the long time of flight, of orders of billions of years, may act as a powerful amplifier of minute quantum gravity effects. There are however also disadvantages: the mechanism of highly energetic particles production at the source is very complicated and not completely understood, while the sources themselves are unpredictable and sparse. 

Complementary to the present and future astrophysical investigations here we would like to put forward the proposal to analyze possible signatures of quantum gravity that could be observed  using current or near future high-energy particle accelerators and sensitive detectors. In contrast to astrophysical phenomena, the  high energy production at accelerators and the subsequent measurements in detectors is characterized  by complete understanding both of the signal and the background, and by huge number of events that can be observed. The apparatus readout and logic of the data acquisition system implemented in the electronics and software, called the {\it trigger} in experimental slang, can be set to select only the sample of events needed in further analysis, thus reinforcing the signal and keeping the background small. 
Insofar as the detector technology is concerned, a spectacular progress in resolutions in time and momentum anticipates pushing the precision frontier by one to two orders of magnitude.
For those reasons it is our working hypothesis that in spite of the considerably lower energies, as compared to astrophysical sources, terrestrial particle accelerators  and detectors can provide a valuable contribution to the quantum gravity phenomenology research program. If this hypothesis is confirmed our investigation would lead to revealing a new field of research that might be capable of providing new significant contribution to the development of the quantum gravity phenomenology research program. In order to address this goal one has to define a class of phenomenologically promising  models, capable of capturing some generic traces of quantum gravitational phenomena and a class of effects for which extremely precise measurements can be performed. Taken together they could single out phenomena of quantum gravitational origin that might be observable at accessible energies many orders of magnitude lower than the Planck scale. 

Before discussing the proposed  experimental setup in details let us first review shortly the current understanding of the theory of $\kappa$-deformed fields. 
They started being investigated shortly after $\kappa$-Poincar\'e algebra was derived  \cite{Lukierski:1992dt}, but it took more than a decade to formulate a complete theory of free $\kappa$-deformed scalar field \cite{Freidel:2007hk}. Recently the free scalar field theory has been studied in depth focusing on the conserved charges associated with global and discrete symmetries \cite{Arzano:2020jro}, \cite{Bevilacqua:2022fbz} and on the  behaviour of multi-particle states under deformed Lorentz transformations \cite{Bevilacqua:2023pqz}. The main upshot of the investigations quoted in the first two papers is that the C, CP, and CPT symmetries are  $\kappa$-deformed, their generators do not commute with boosts. From this it follows that although a particle and antiparticle have the same rest masses, after the boost their kinematic characteristics may differ. The analysis of massless and massive scalar field is to be shortly followed by investigations of fields of higher spins.

One of the straightforward phenomenological motivations for extending the $\kappa$-deformed formalism to massive vector and fermion fields is the accuracy challenge provided by the anomalous magnetic moment of the muon, commonly known as the ``$g-2$'' problem \cite{aoyama}.
Current studies focus on small but significant discrepancies between the experimental data and calculations reaching an accuracy of eleven significant digits. 
Since part of the effort is now focused on the determination of loop corrections to the vacuum polarization, it seems important to investigate the sensitivity of these corrections to the quantum gravitational effects (in the guise of $\kappa$-deformation) in the photon propagator.

\vskip 5mm

\begin{tabular}{cc}
\hskip 1cm \includegraphics[scale=0.25]{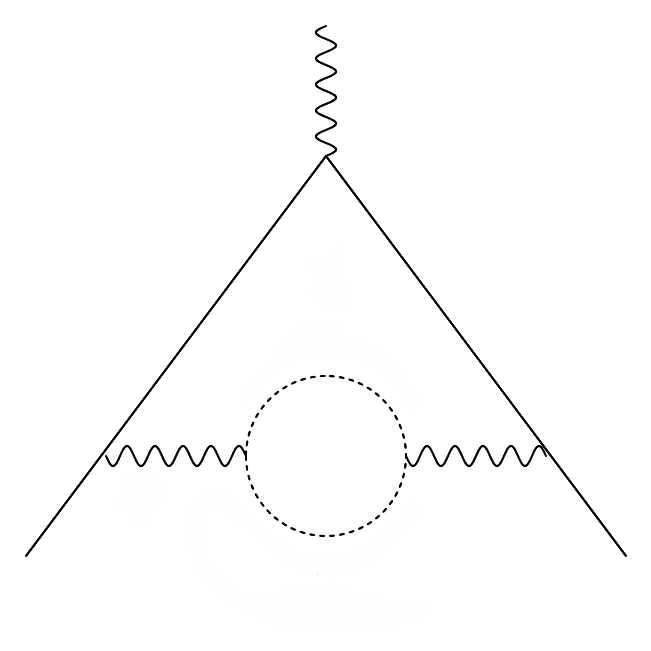} & \hskip 15mm \includegraphics[scale=0.22]{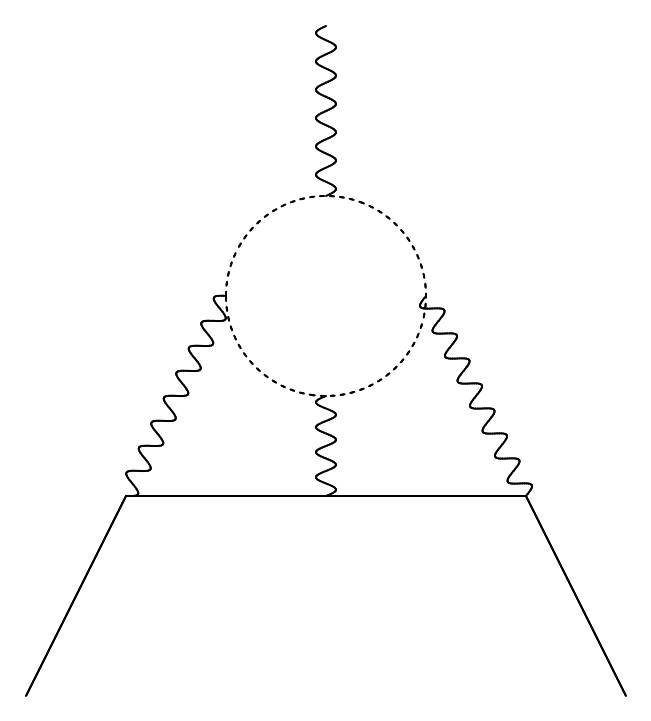}
\end{tabular}
\begin{center}
    \textbf{Figure 1}
\end{center}

\vskip 5mm

Example of loop processes contributing to the anomalous magnetic moment of the muon are depicted in the figure above where the left diagram shows the hadronic vacuum polarization and the right one is called the light-by-light scattering. 
The four-momentum composition at all vertices gets modified and is performed using the deformed summation of momenta $p\oplus q$ and  minus (antipode) $S(p)$, as defined in Ref. \cite{akw_plb2019}. 
\newline



Another possible observational opportunities might be connected with deformations of discrete symmetries. In $\kappa$-deformed field theory the charge-conjugation operator C
does not change the sign of the quantum numbers of particles and does not commute with boosts. As a consequence even if the particle and antiparticle have the same rest mass, when acting on a particle state of given  the charge conjugation operator C produces an antiparticle state of slightly different momentum \cite{Bevilacqua:2022fbz}.
Thus the combined symmetry transformations CP and CPT get $\kappa$-deformed in the Lorentz-boosted frames.

In experimental search for possible deviations from the CP and CPT symmetries one makes use of the observables called the asymmetries of decay rates.
Denoting by $|P\ra$ an initial state of an unstable particle, and by $|f\ra$ the final state being the product of its decay, and using a bar sign to denote the corresponding antiparticle's states, let us define the decay amplitudes 
\begin{eqnarray}
\label{amplit}
  A_f  = \la f|H|P\ra & \quad A_{\bar f} = \la \bar f|H|P\ra \nn \\
  \bar A_f  =  \la f|H|\bar P\ra & \quad \bar A_{\bar f} = \la \bar f|H|\bar P\ra.
\end{eqnarray}
Using these  amplitudes and their symmetry properties one defines two types of asymmetries sensitive to the CP and CPT violations
\be
\label{asym}
{\cal A_{CP}} & = & \frac{|\bar A_f|^2-|A_{\bar f}|^2}{|\bar A_f|^2+|A_{\bar f}|^2} \nn \\
{\cal A_{CPT}} & = & \frac{|\bar A_{\bar f}|^2-|A_{f}|^2}{|\bar A_{\bar f}|^2+|A_{f}|^2},
\ee
where both asymmetries in Eq. (\ref{asym}) can depend on decay times of particles.

The initial states $|P\ra$ and $|\bar P\ra$ in Eq. (\ref{amplit}) are flavor eigenstates.
In order to follow their time evolution, governed by the Hamiltonian, one needs to express them in the basis of energy eigenstates.
For example, for $P$ representing a heavy, neutral pseudoscalar mesons ($K^0$, $D^0$, $B^0$,..), such decomposition reads
\be
\label{pdef}
|P\ra & = & \frac{1}{\sqrt{2}}(|P_L\ra + |P_H\ra) \nn \\
|\bar P\ra & = & \frac{1}{\sqrt{2}}(|P_L\ra - |P_H\ra),
\ee
where parameters quantifying CP violation are omitted for simplicity \footnote{In the fully-fledged calculations, the CP violation is accounted for by an efficient parametrization using the mixing and direct CP violation parameters $\varepsilon\simeq 10^{-3}$ and $\varepsilon^\prime \simeq 10^{-6}$ weighting linearly the states (\ref{pdef})}.

\vskip 5mm

\begin{tabular}{cc}
\hskip 1cm \includegraphics[scale=0.25]{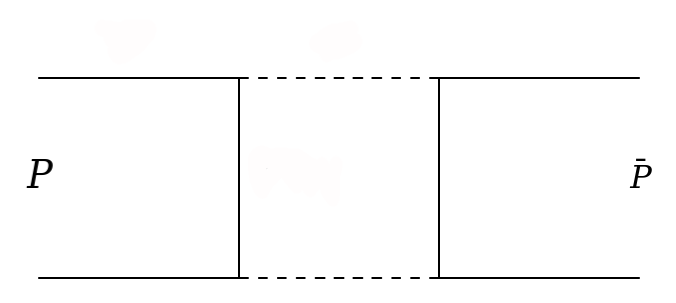} & \hskip 15mm \includegraphics[scale=0.22]{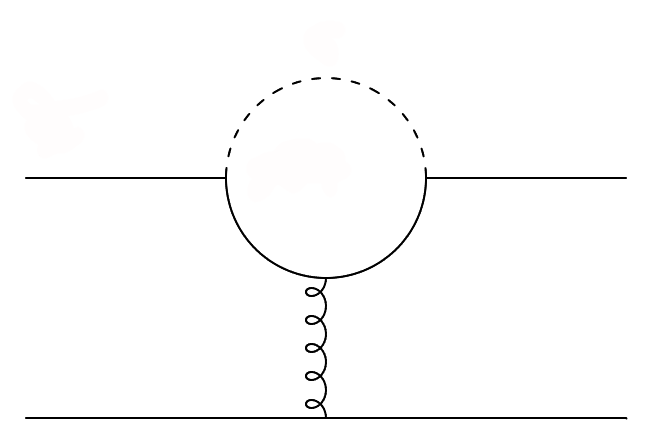}\\
\hskip 1cm (a) & \hskip 15mm (b)
\end{tabular}
\begin{center}
    \textbf{Figure 2}
\end{center}

\vskip 5mm

The processes responsible for CP violation are, as depicted in the diagrams above, (a) the mixing where the flavour $F$ of the meson changes by two units, $\Delta F=2$  and (b) the direct CP violation where flavour changes by one unit, $\Delta F=1$, where $F$ stands for either the strangeness, charm or beauty flavour.

Similarly to the vacuum polarization diagrams contributing to the anomalous magnetic moment of the muon, the four-momentum balance in CP-violating amplitudes get modified and the deformed summation $p\oplus q$ is used instead of normal sum $p+q$ \cite{akw_plb2019}.

As shown in Ref. \cite{Bevilacqua:2022fbz}, deformation of the charge conjugation operator C leads to the dependence of the particle state $|P(p)\ra$ on the four-momentum $p$ becomes modified for its antiparticle $|\bar P(S(p))\ra$, where 
\be
\label{Sp}
S(p)_0\simeq                    -E+\frac{{\bf p}^2}\kappa \,,\quad \mathbf{S}  (p)\simeq    -{\bf p}   +      {\bf p}\,  \frac{E}\kappa
\ee
This leads to a different time evolutions of the amplitudes (\ref{amplit})
\be
\label{amplitudes}
A_f & = & \frac{1}{\sqrt{2}} [\la f|P_L\ra e^{-iE_Lt_L-(\Gamma_L/2)t_L}+\la f|P_H\ra e^{-iE_Ht_H-(\Gamma_H/2)t_H}] \nn \\
\bar A_f & = &  \frac{1}{\sqrt{2}}[\la f|P_L\ra e^{-i(E_L-\delta)t_L-(\Gamma_L/2)t_L}-\la f|P_H\ra e^{-i(E_H-\delta)t_H-(\Gamma_H/2)t_H}] \nn \\
A_{\bar f} & = &  \frac{1}{\sqrt{2}}[\la \bar f|P_L\ra e^{-iE_Lt_L-(\Gamma_L/2)t_L}+\la\bar f|P_H\ra e^{-iE_Ht_H-(\Gamma_H/2)t_H}] \nn \\
\bar A_{\bar f} & = &  \frac{1}{\sqrt{2}}[\la \bar f|P_L\ra e^{-i(E_L-\delta)t_L-(\Gamma_L/2)t_L}-\la\bar f|P_H\ra e^{-i(E_H-\delta)t_H-(\Gamma_H/2)t_H}],
\ee
where $E_{L,H}, \Gamma_{L,H}$ are the energy and decay width eigenvalues of the light and heavy eigenstates, $t_{L,H}$ - their evolution time, and $\delta={\bf p}^2/\kappa$ stands for a correction dependent on deformation $\kappa$.
This modification  is non-zero only in a reference frame where the particle has a non-zero momentum.
For a particle being the decay product of a heavier unstable particle, the $\delta$ correction can be non-zero in the parent's centre-of-momentum frame.
However, this may not be enough to make the effect sizeble and therefore one needs a strong Lorentz boost to a fast moving reference frame.

This correction affects both the intensity and oscillation frequency in an interference term in dependence of asymmetries on the particle decay time (\ref{asym}).
After integration over time it remains in final expressions and therefore can be detected also in time-integrated asymmetries.
Similar effects can be found for the CPT asymmetries ${\cal A}_{CPT}$ (\ref{asym}), both the time-dependent and time-integrated ones.
\newline

The four-momentum dependence of deformation under C operation has also intriguing consequences for interference of fields.
The cleanest example comes from interference of flavoured mesons produced in decays of vector resonances $J^{PC}=1^{--}$ into pseudoscalars, e.g.:
\be
\label{channels}
\phi^0 (1020) & \rightarrow & K^0 \bar K^0 \nn \\
\Psi(3770) & \rightarrow & D^0\bar D^0 \nn \\
\Upsilon(10580) & \rightarrow & B^0\bar B^0 \nn \\
\Upsilon(10860) & \rightarrow & B^0_s\bar B^0_s
\ee
The final states in channels (\ref{channels}) are coherent and they exhibit quantum-mechanical entanglement.

In normal quantum mechanics, decay channel of $\phi^0$ was analysed in details and for some, the most frequent, final states of decaying kaons also experimental studies have been performed (cf. Refs. \cite{kloe_book,kloe_1}). 

The two-particle states (\ref{channels}) $|\psi\ra$ must be antisymmetric, which means that the states composed from the meson and antimeson $|P\ra|\bar P\ra$ and $|\bar P\ra|P\ra$ contribute with opposite signs.
Keeping in mind that the antiparticle state is deformed (\ref{Sp}), $|\psi\ra$ reads
\be
\label{psi}
|\psi\ra = & \frac{1}{\sqrt{2}}(|P(p)\ra|\bar P(S(p))\ra-|\bar P(-S(p))\ra|P(-p)\ra).
\ee

The two-particle decay-time spectrum exhibits a characteristic oscillatory pattern due to interference of the mass eigenstates $|P_H\ra$ and $|P_L\ra$.
The interference process is described by a box-like diagram in Figure 2, where, again, the four-momentum balances in vertices are deformed \cite{akw_plb2019}.

However, because of the deformation of antiparticle's fields (\ref{Sp}), in the $(L,H)$ basis, in addition to the states with the correct CPT parity $|P_L\ra|P_H\ra, |P_H\ra|P_L\ra$ the states with a "wrong" CPT parity: $|P_L\ra|P_L\ra, |P_H\ra|P_H\ra$ are also present.
Physically, this means that incorporation of gravity may blur the concept of antiparticle. 
The emergence of this kind of states leads to a violation of CPT and was introduced ad hoc in the model of Ref. \cite{bernabeu_1}.
Here we propose how to derive, and possibly observe, these states extending from the more fundamental, quantum-gravitational concepts.
Consequently, the state (\ref{psi}) is not CPT-invariant due to the Lorentz-frame dependent and $\kappa$-dependent corrections.

The decay intensity $|\la f_1,f_2|T|\psi\ra|^2$ consists of three parts: the correct-parity term, wrong-parity term, and interference between those two.
Each of them gets modified in both the amplitude and oscillation frequency.

In addition to effects on the interference pattern described above, modification of kinematics of the two-particle states is expected with no interference.
These effects were studied in Ref. \cite{Bevilacqua:2023pqz} and need to be quantitatively estimated in the experimental context.
It is common to study interference patterns in variables defined as a difference of particles’ momenta or difference of their decay times. Sensitivity of these variables to new phenomena is the best for their small values where interference pattern variates largely and statistics of data samples is sizable.
Modifications of angular distributions in a boosted frame can be also detectable using very sensitive detectors.
\newline

Phenomenological predictions for deformed fields strain to isolate extremely subtle effects of deformation of four momenta, leading to modification of the Lorentz boost, typically by $\sim p^2/(m\kappa)$, where $m\sim 1$ GeV and $\kappa$ is of order of the Planck scale $10^{19}$ GeV.
As shown in Refs. \cite{akw_plb2019, Bevilacqua:2022fbz}, one may expect to set measurable limits from lifetimes of particles vs. antiparticles of few orders of magnitudes below the Planck mass, e.g $\kappa > 10^{14}$ GeV at LHC and even further to $\kappa> 10^{16}$ GeV using the future collider \cite{fcc}.
However, in practice such a measurement requires a long baseline experiment with accuracy at the limit of current technology.

Our recent research indicates a more promising measurement being that of the oscillation frequency in the decay time difference of interfering particles.
Both the correct-CPT and wrong-CPT terms get a $p^2/\kappa$ correction to the oscillation frequency $\Delta E$, providing better sensitivity compared to the amplitude modification, in terms $\cos (\Delta E+p^2/\kappa)\Delta t$.
Analogies to the amplitude and frequency modulations in radiotechnology provide a well-justified hope.

Our preliminary studies indicate that the strength of predictions leading to estimates of $\kappa$ in real experiments lies in the accuracy of detectors.
Accuracy of the particle lifetime measurements in high-energy experiments critically depends on the spatial and time resolutions of detectors and combination of data when decay events are reconstructed.
One of the best time resolution, provided by the LHCb detector at CERN \cite{lhcb_det}, amounts to 45 fs and this figure has been used for our estimates in Ref. \cite{Bevilacqua:2022fbz}. 
Possible progress in detection techniques
leading to better time resolution can move the limits on $\kappa$
even further. 
Particularly interesting would be imaging
technologies based on femtosecond lasers that could hopefully improve time resolutions to be around 1 fs and below \cite{hongwei}.
A lot of research in this direction has been performed and the recommendations are being continuously given in the recommendations of the European Committee for Future Accelerators and the project Advancement and Innovation for Detectors at Accelerators run at CERN \cite{ecfa}.

Quantum-gravitational corrections to many phenomenological quantities such as the mesons' oscillation frequencies, resonance lifetimes, violation of discrete symmetries, and contributions to the muon's anomalous magnetic moment, will be most likely overshadowed by the uncertainties of parameters quantifying them in experiments. For instance, the CP violation parameters are best known for neutral kaons but the errors on the $\varepsilon$ and $\varepsilon^\prime$ parameters, quantifying the indirect and direct CP violation, are $10^{-5}$ and $10^{-7}$, respectively. 
Even if one succeeds to calculate the $\kappa$-deformed diagrams considered in this project, their contribution  could be smaller than other uncertainties. This project, however, aims at proposing the gravity-sensitive observables and new requirements to detector technologies, stringent enough to measure them. 

To conclude: we propose to undertake the research project aimed at  finding the processes and observables that are most sensitive to the deformations of quantum gravitational origin and to set the quantitative challenges to experimental technologies. This project will result in new methods to detect traces of quantum-gravity induced deformations of discrete symmetries.


\begin{thebibliography}{99}
\bibitem{Bronstein:2012zz}
M.~Bronstein,
``Quantum theory of weak gravitational fields,''
Gen. Rel. Grav. \textbf{44} (2012), 267-283, reprint 

\bibitem{Dyson:2013hbl}
F.~Dyson,
``Is a graviton detectable?,''
Int. J. Mod. Phys. A \textbf{28} (2013), 1330041

\bibitem{Amelino-Camelia:1999hpv}
G.~Amelino-Camelia,
``Are we at the dawn of quantum gravity phenomenology?,''
Lect. Notes Phys. \textbf{541} (2000), 1-49
[arXiv:gr-qc/9910089 [gr-qc]].

\bibitem{Amelino-Camelia:2008aez}
G.~Amelino-Camelia,
``Quantum-Spacetime Phenomenology,''
Living Rev. Rel. \textbf{16} (2013), 5
[arXiv:0806.0339 [gr-qc]].

\bibitem{Addazi:2021xuf}
A.~Addazi, J.~Alvarez-Muniz, R.~Alves Batista, G.~Amelino-Camelia, V.~Antonelli, M.~Arzano, M.~Asorey, J.~L.~Atteia, S.~Bahamonde and F.~Bajardi, \textit{et al.}
``Quantum gravity phenomenology at the dawn of the multi-messenger era\textemdash{}A review,''
Prog. Part. Nucl. Phys. \textbf{125} (2022), 103948
[arXiv:2111.05659 [hep-ph]].



\bibitem{Amelino-Camelia:1997ieq}
G.~Amelino-Camelia, J.~R.~Ellis, N.~E.~Mavromatos, D.~V.~Nanopoulos and S.~Sarkar,
``Tests of quantum gravity from observations of gamma-ray bursts,''
Nature \textbf{393} (1998), 763-765
[arXiv:astro-ph/9712103 [astro-ph]].

\bibitem{cost} Project CA18108 ``Quantum Gravity Phenomenology In The Multi-Messenger
Approach'', https://www.cost.eu/actions/CA18108, and https://qg-mm.unizar.es/

\bibitem{Hossenfelder:2012jw}
S.~Hossenfelder,
``Minimal Length Scale Scenarios for Quantum Gravity,''
Living Rev. Rel. \textbf{16} (2013), 2
[arXiv:1203.6191 [gr-qc]].

\bibitem{Doplicher:1994zv}
S.~Doplicher, K.~Fredenhagen and J.~E.~Roberts,
``Space-time quantization induced by classical gravity,''
Phys. Lett. B \textbf{331} (1994), 39-44

\bibitem{Doplicher:1994tu}
S.~Doplicher, K.~Fredenhagen and J.~E.~Roberts,
``The Quantum structure of space-time at the Planck scale and quantum fields,''
Commun. Math. Phys. \textbf{172} (1995), 187-220
[arXiv:hep-th/0303037 [hep-th]].

\bibitem{Majid:1999tc}
S.~Majid,
``Meaning of noncommutative geometry and the Planck scale quantum group,''
Lect. Notes Phys. \textbf{541} (2000), 227-276
[arXiv:hep-th/0006166 [hep-th]].

\bibitem{Lukierski:1991pn}
J.~Lukierski, H.~Ruegg, A.~Nowicki and V.~N.~Tolstoi,
``Q deformation of Poincare algebra,''
Phys. Lett. B \textbf{264} (1991), 331-338

\bibitem{Lukierski:1991ff}
J.~Lukierski, A.~Nowicki and H.~Ruegg,
``Real forms of complex quantum anti-De Sitter algebra U-q(Sp(4:C)) and their contraction schemes,''
Phys. Lett. B \textbf{271} (1991), 321-328
doi:10.1016/0370-2693(91)90094-7
[arXiv:hep-th/9108018 [hep-th]].

\bibitem{Lukierski:1992dt}
J.~Lukierski, A.~Nowicki and H.~Ruegg,
``New quantum Poincare algebra and k deformed field theory,''
Phys. Lett. B \textbf{293} (1992), 344-352

\bibitem{Lukierski:1993wx}
J.~Lukierski, H.~Ruegg and W.~J.~Zakrzewski,
``Classical quantum mechanics of free kappa relativistic systems,''
Annals Phys. \textbf{243} (1995), 90-116
[arXiv:hep-th/9312153 [hep-th]].

\bibitem{Majid:1994cy}
S.~Majid and H.~Ruegg,
``Bicrossproduct structure of kappa Poincare group and noncommutative geometry,''
Phys. Lett. B \textbf{334} (1994), 348-354
[arXiv:hep-th/9405107 [hep-th]].

\bibitem{Arzano:2021scz}
M.~Arzano and J.~Kowalski-Glikman,
``Deformations of Spacetime Symmetries: Gravity, Group-Valued Momenta, and Non-Commutative Fields,'', Lecture Notes in Physics, Springer, 2021.


\bibitem{Amelino-Camelia:2000cpa}
G.~Amelino-Camelia,
``Testable scenario for relativity with minimum length,''
Phys. Lett. B \textbf{510} (2001), 255-263
[arXiv:hep-th/0012238 [hep-th]].

\bibitem{Amelino-Camelia:2000stu}
G.~Amelino-Camelia,
``Relativity in space-times with short distance structure governed by an observer independent (Planckian) length scale,''
Int. J. Mod. Phys. D \textbf{11} (2002), 35-60
[arXiv:gr-qc/0012051 [gr-qc]].

\bibitem{Amelino-Camelia:2011lvm}
G.~Amelino-Camelia, L.~Freidel, J.~Kowalski-Glikman and L.~Smolin,
``The principle of relative locality,''
Phys. Rev. D \textbf{84} (2011), 084010
[arXiv:1101.0931 [hep-th]].

\bibitem{Freidel:2007hk}
L.~Freidel, J.~Kowalski-Glikman and S.~Nowak,
``Field theory on kappa-Minkowski space revisited: Noether charges and breaking of Lorentz symmetry,''
Int. J. Mod. Phys. A \textbf{23} (2008), 2687-2718
[arXiv:0706.3658 [hep-th]].

\bibitem{Arzano:2020jro}
M.~Arzano, A.~Bevilacqua, J.~Kowalski-Glikman, G.~Rosati and J.~Unger,
``$\kappa$-deformed complex fields and discrete symmetries,''
Phys. Rev. D \textbf{103} (2021) no.10, 106015
[arXiv:2011.09188 [hep-th]].

\bibitem{Bevilacqua:2022fbz}
A.~Bevilacqua, J.~Kowalski-Glikman and W.~Wislicki,
``\ensuremath{\kappa}-deformed complex scalar field: Conserved charges, symmetries, and their impact on physical observables,''
Phys. Rev. D \textbf{105} (2022) no.10, 105004
[arXiv:2201.10191 [hep-th]].

\bibitem{Bevilacqua:2023pqz}
A.~Bevilacqua, J.~Kowalski-Glikman and W.~Wi\'slicki,
[arXiv:2305.09180 [hep-th]].

\bibitem{Hersent:2022gry}
K.~Hersent, P.~Mathieu and J.~C.~Wallet,
``Gauge theories on quantum spaces,''
Phys. Rept. \textbf{1014} (2023), 1-83
[arXiv:2210.11890 [hep-th]].

\bibitem{aoyama} T. Aoyama et al., ''The anomalous magnetic moment of the muon in the Standard Model'', Phys. Rept. \textbf{887} (2020) 1-166

\bibitem{akw_plb2019} A. Arzano, J. Kowalski-Glikman and W. Wiślicki, "A bound on Planck-scale deformations of CPT from muon lifetime", Phys. Lett. {\bf B794} (2022) 41

\bibitem{pdg} R.L. Workman et al. (Particle Data Group), "The Review of Particle Physics  (2022)", Prog. Theor. Exp. Phys. 2022, 083C01 (2022), sections ``CP violation in the quark sector'', ``CPT Invariance Tests in Neutral Kaon Decay'', ``$D^0 - \bar D^0$ mixing'', ``$B^0 - \bar B^0$ mixing''

\bibitem{branco} G.C. Branco, L. Lavoura and J.P. Silva, "CP Violation", Clarendon Press, Oxford, 1999

\bibitem{kostelecky} V.A. Kostelecky, "CPT, T, and Lorentz violation in neutral-meson oscillations", Phys. Rev. {\bf D64} (2001) 076001

\bibitem{kloe_book} ``Handbook on neutral kaon interferometry at a $\phi$-factory'', ed. A. di Domenico, Frascati Physics Series, Frascati 2007

\bibitem{fcc} https://home.cern/science/accelerators/future-circular-collider.

\bibitem{kloe_1} KLOE-2 Collaboration, D. Babusci et al., "Precision tests of quantum mechanics and CPT symmetry with entangled neutral kaons at KLOE", JHEP \textbf{04} (2022) 059 

\bibitem{bernabeu_1} J. Bernabeu, N.E. Mavromatos and J. Papavassiliou, "Novel type of CPT violation for correlated EPR states", Phys.Rev.Lett. \textbf{92} (2004) 131601

\bibitem{lhcb_det} LHCb Collaboration, R. Aaij et al., ''LHCb detector
performance'', Int. J. Mod. Phys. A \textbf{30}, 1530022 (2015)

\bibitem{hongwei} Hongwei Chen, ''Toward unlimited temporal resolution:
femtosecond videography for atomic and molecular dynamics, Light Sci. Appl. \textbf{6}, e17123 (2017); https://www.nature.com/articles/lsa2017123.

\bibitem{ecfa} https://aidainnova.web.cern.ch/update-ecfa-roadmap-rd-detector-technologies


\bibitem{Hersent:2022gry}
K.~Hersent, P.~Mathieu and J.~C.~Wallet,
``Gauge theories on quantum spaces,''
Phys. Rept. \textbf{1014} (2023), 1-83
doi:10.1016/j.physrep.2023.03.002
[arXiv:2210.11890 [hep-th]].



























\end{thebibliography}
\end{document}